\documentclass[12pt]{article}
\def\be{\begin{equation}}
\def\ee{\end{equation}}
\def\la{\label}
\def\bea{\begin{eqnarray}}
\def\eea{\end{eqnarray}}

\def\la{\label}
\def\bib{\bibitem}

\def\Lm{\Lambda}
\def\le{\left}
\def\ri{\right}

\def\al{\alpha}

\def\s8{\sigma_8}

\def\fr{\frac}

\usepackage{graphicx}

\begin{document}

\begin{center}
   {\Large \bf  Inflation at the maxima of symmetric potentials}

\end{center}

\vspace*{0.5cm}

\begin{center}
{\bf G. Germ\'an$^a$\footnote{e-mail: gabriel@fis.unam.mx} and  A.
de la Macorra$^b$\footnote{e-mail: macorra@fisica.unam.mx} }
\end{center}


\begin{center}
{\small
\begin{tabular}{c}
$^a$ Centro de Ciencias F\'{\i}sicas, UNAM \\
Apdo. Postal 48-3, 62251 Cuernavaca, Morelos,  M{\'e}xico\\
$^b$ Instituto de F\'{\i}sica, UNAM\\
Apdo. Postal 20-364, 01000  M\'exico D.F., M\'exico\\

\\
\end{tabular}
}
\end{center}


\begin{center}
{\bf ABSTRACT}
\end{center}
\small{ We construct a two-stage inflationary model which can
accommodate early inflation at a scale $\Lambda_1$ as well as a
second stage of inflation at $\Lambda_2$ with a
single scalar field $\phi$.
  We use a symmetric potential, valid in a frictionless world,
in which the two inflationary periods have exactly the same scale,
i.e. $\Lambda_1=\Lambda_2$. However, we see today $\Lambda_1 \gg
\Lambda_2$ due to the friction terms (expansion of the universe
and interaction with matter). These type of models can be
motivated from supergravity.
  Inflation occurs close to the maxima of the potential. As a
consequence both inflations are necessarily finite. This opens the
interesting possibility that the second inflation has already or
is about to end.  A first inflation is produced when fluctuations
displace the inflaton field from its higher maximum rolling down
the potential as in new inflation. Instead of rolling towards a
global minimum the inflaton approaches a lower maximum where a
second inflation takes place. }

 \vspace*{.1 cm}

\noindent \rule[.1in]{14.5cm}{.002in}

\thispagestyle{empty}

\setcounter{page}{0} \vfill\eject

\section{Introduction} \label{intro}

The idea that the universe underwent an early inflationary
expansion is now widely accepted \cite {reviews}. This era of
inflation makes plausible certain initial conditions for standard
cosmology and provides a mechanism for structure formation. More
speculatively the idea that the universe is at present undergoing
inflation (usually denoted by the term quintessence) is the
subject of much current interest \cite {papers}. Several models
have been proposed where typically the potential energy of a
scalar field, in general different from the one producing early
inflation, is dominating the dynamics of the universe. Usually the
potential is an inverse power of the field decreasing
monotonically towards zero. In the present work we are interested
in studying a model which accommodates two stages of inflation by
the evolution of a single scalar field \cite {vilenkin}. Here,
however, we look at the possibility that both inflations are
produced when the inflaton is close to the maxima of the
potential. The fact that both inflations occur at the maxima
implies that they are necessarily finite. This opens the
interesting possibility where the second inflation has already or
is about to end. This possibility is not ruled out by existing
data and could be testable with far more, higher accuracy,
supernovae on the Hubble diagram \cite {filippenko}.

We take the position that inflation occurs not only at the maxima
of the potential but also that these two stages of inflation, the
initial and the present day, have the same value $V(M_1)=V(M_2)$.
Where $M_1$ and $M_2$ denote the first and second maxima of the
potential, respectively. The reason why we see today
$V(M_1) \gg V(M_2)$ is due to the existence of friction terms. The
friction terms are given by the expansion of the universe and by
the interaction of the inflaton field with matter.

In what follows we develop this idea by using an analogy with a
problem from classical mechanics. We discuss a simple toy model
illustrating the main points. The resulting potential, when
rewritten for the inflaton, could also be obtained from supergravity
(see Appendix).

\begin{figure}[htp!]
\begin{center}
\includegraphics[width=6cm]{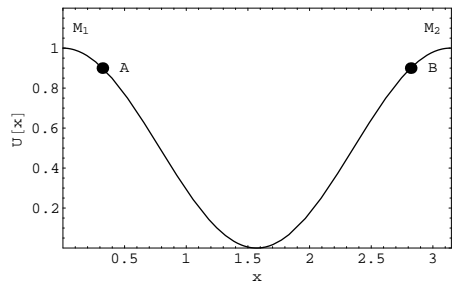}
\end{center}
\caption{\small{The potential  energy $U(x)$ of a particle in
classical mechanics. In the absence of friction if we leave the
particle at the point $A$ with vanishing velocity it will
eventually reach $B$, with $U(A)=U(B)$, also with vanishing velocity.
In the limit when $A \rightarrow M_1$ it will take an infinite
amount of time for the particle to reach $B \rightarrow M_2$.}}
\la{fig1}
\end{figure}

Let us consider a potential $U(x)$ as shown in Fig. \ref{fig1}.
When there is no friction the equation of motion for a particle of
mass $m=1$ is given by
\begin{equation}
\ddot{x}+U'(x)=0
\label{newton}
\end{equation}
where $\dot x\equiv dx/dt$ and $U'\equiv dU/dx$. The conserved
quantity of eq.(\ref{newton}) is just the energy $E=\dot x^2/2+
U(x)$. We study the problem of a particle that leaves with
vanishing velocity somewhere from the left of the minimum, let us
say $A$ and reaches a maximum height $B$ some time later with
$U(A)=U(B)=E$. If we fix the origin of time at the minimum of
$U(x)$ then the particle leaves $A$ in the past reaching $B$
sometime in the future. As $A$ becomes close to the maximum at
$M_1$ the particle spends longer close to the maxima. In the limit
when $A \rightarrow M_1$ it takes an infinite amount of time for
the particle to reach $M_2$. The particle would spend most of the
time leaving $M_1$ and trying to reach $M_2$. As a result the
kinetic energy is negligible close to the maxima; the potential
energy dominates. We call this the limiting solution. The maximum
at $M_1$ is located at $x=0$ thus we require $x(t=-\infty)=0$ and
$x(t=+\infty)$ locates the maximum at $M_2$. As a concrete example
let us consider the potential
\begin{equation}
U(x)=E \cos^2(x).
\label{pot1}
\end{equation}
It is easy to check that the limiting solution is
\begin{equation}
x(t)=(2 \arctan[\tanh(t\sqrt{\frac{E}{2}})]+\frac{\pi}{2}),
\label{sol1}
\end{equation}
where $x(t=-\infty)=0$, $x(t=+\infty)=\pi$ and
$\dot{x}(t=-\infty)=\dot{x}(t=+\infty)=0$.
The potential $U(x)$ is already illustrated in Fig. \ref{fig1}.
 If we could lower the r.h.s.
 branch of this potential we
could use this mechanical problem as an analogy to construct a
model with two stages of inflation. Actually this can be done as
follows. Instead of the symmetric potential $U(x)$ let us consider
a new potential $V(x)$, which we call the asymmetric potential,
illustrated in Fig. \ref{fig2}. Now the maximum at $M_2$ is much
smaller than the maximum at $M_1$. It is clear that we need a
friction term in the corresponding Eq.(\ref{newton})  to stop the
particle precisely at $M_2$. Instead of eq.(\ref{newton}) we would
have
\be\la{newfric}
\ddot x +c \dot x + V'(x)=0.
 \ee
Eq.(\ref{newfric}) is similar to eq.(\ref{newton}) if we replace $U(x)$ by
the "potential" $U'(t)=V'(x) + c \dot x$. Furthermore,
eq.(\ref{newfric}) gives a conserved quantity $E=\dot x^2/2 +
U(t)$ where $U(t)=V(x) +\int_{x_i}^{x} c \dot x dx=V(x)
+\int_{t_i}^{t} c \dot x^2 dt$ since $\dot E=\dot x(\ddot x +c
\dot x + V'(x))=0$. Notice that $E$ is in general not the energy
since at a given point (t,x(t)) it depends on the history of the
trajectory trough the integral $\int_{t_i}^{t} c \dot x^2 dt$ in $U(t)$
and only in the case $c=0$ (i.e. no friction) $E$ is the conserved
energy. However, since $E$ is conserved even for $c \neq 0$ one has
a maximum of the  potential $U(t_i)=U(t_f)$ at $\dot x(t_i)=\dot x(t_f)=0$,
i.e. the height of the potential is the same, and
the two maximum values of the potential $V(x)$  are then given by
\be\la{max}
V(x_i)=U(t_i) \hspace{1cm} V(x_f)=U(t_f)-\int_{x_i}^{x_f} c \dot x
dx.
\ee
For $\int_{x_i}^{x_f} c \dot x dx > 0$ we can easily
have $V(x_f) \ll V(x_i)=U(t_i)=U(t_f)$. Using the conservation of
$E$ we can write the friction term in the following equivalent
forms \be\la{fric} \int_{x_i}^{x} c \dot x dx=  \int_{t_i}^{t} c
\dot x^2 dt=\int_{t_i}^{t} c 2(E-U) dt. \ee
If we use the potential in eq.(\ref{pot1}) we can
integrate the friction term giving $\int_{x_i}^{x_f} c \dot x dx=
c\sqrt{2E} (cos(x_i)-cos(x_f))$ and taking $x_i=0, x_f=\pi$ the
potential at the second maximum is
$V(x_f)=U(t_f)-2c\sqrt{2E}=E-2c\sqrt{2E}$ which is smaller than
$V(x_i)=E$. The  limiting solution eq.(\ref{sol1}) solves
eqs.(\ref{newton}) and (\ref{newfric}) and imposing this solution
to the potential  $V(x)$ determines the friction term or
equivalently if we know the friction term we can determine the
energy scale at the maxima with $M_2 \ll M_1$ where the particle
stops.

\begin{figure}[htp!]
\begin{center}
\includegraphics[width=6cm]{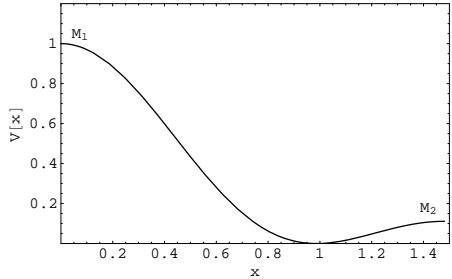}
\end{center}
\caption{\small{A particle leaves the  maximum at $M_1$ with
vanishing velocity. It will just reach $M_2$ also with vanishing
velocity if there is a friction term which stops the particle
precisely at $M_2$.}} \la{fig2}
\end{figure}

In inflationary models of the "new" type one typically starts with
a very flat potential and inflation occurs close to the maximum at
$\phi=0$, where $\phi$ is the inflaton field. There could be a
previous "primordial" stage of inflation probably of the chaotic
type setting the initial conditions for new inflation. For
simplicity in what follows we will call this new inflationary
epoch a first stage or simply first inflation   characterized by a
scale $\Lambda_1$. This scenario is illustrated in Fig. \ref{fig2}
with $x(t) \rightarrow \phi(t)$. Here we study the possibility of
a second stage of inflation at a scale $\Lambda_2 (t_2)$, where
$\Lambda_2(t_2)\ll\Lambda_1(t_1)$. The mechanical analogy
indicates that the second inflation will occur also close to a
maximum.

The value of the potential at the second inflationary period $t_2$
from  a symmetric point of view is exactly the same as that of the
first inflationary epoch at $t_1$, i.e. $U(t_1)=U(t_2)$, but
because we live in a asymmetric world (a world with friction
terms) we see that $V(t_2)=\Lambda_2^4 \ll V(t_1)=\Lambda_1^4$.
We propose that it is only due to the dynamics including friction
terms that we are now seeing a very small inflationary scale
compared to the first one.

\section{The Cosmological Model} \label{model}

The inflaton field equation of motion
and  Friedman's equation are, as usual, given
by
\begin{equation}
\ddot{\phi}+3H\dot{\phi}+V'(\phi)=0, \label{fi}
\end{equation}
\begin{equation}
3H^2=\rho,
\label{fridmann}
\end{equation}
where we have set the reduced Planck mass $M=2.44 \times 10^{18}
GeV$ to unity and $\rho$ is the total energy density and we are
considering a flat universe and an homogenous scalar inflaton field.
In analogy with eq.(\ref{newfric}) we will assume a symmetric potential
$U(t)$ such that $U(t_i)=U(t_f)$ at $\dot\phi(t_i)=\dot\phi(t_f)==0$.
>From eq. (\ref{fi}) and
$U'(t)\equiv 3H\dot{\phi}+V'(\phi)  $ one has
\be\la{vv0}
U(t)=V(\phi) +\int_{\phi_i}^{\phi}3H\dot{\phi} d\phi=V(\phi)
+\int_{t_i}^{t}3H\dot{\phi}^2 dt
\ee
where the integration constant has been fixed by demanding
$V_i=U_i$ and we have used $d\phi=\dot\phi dt$. From eq.(\ref{fi})
we can define a conserved quantity $E_\phi=\dot \phi^2/2+U(t)$,
with $\dot E_\phi=\dot\phi(\ddot{\phi}+3H\dot{\phi}+V'(\phi))=0$,
and at the points $\dot\phi(t_a)=0$ one has the maximum of the potential
$U(t_a)$. The derivative of $H$ is
\begin{equation}
\dot{H}=-\frac{1}{2}(\dot{\phi}^2+\rho_a(1+w_a))
\label{hdot}
\end{equation}
where we have assumed $\rho=\rho_\phi+\rho_a$ with
$\rho_\phi=\dot \phi^2/2+V(\phi)$ and $\rho_a$ is the energy
density of matter or radiation. From
eq.(\ref{hdot}) the friction term  gives
\be\la{fricH}
\int_{\phi_i}^{\phi}3H\dot{\phi}d\phi=-\int_{t_i}^{t}3(2H\dot
H+H\rho_a(1+w_a)) dt =3\Delta H^2-\Delta \rho_a
\ee
with $dH^2/dt=2H\dot H$, $\dot\rho_a=-3H\rho_a(1+w_a)$ and $\Delta
H^2=H^2_i-H^2(t)\geq 0, \Delta \rho_a=\rho_{a i} -\rho_{a}(t)\geq
0$. At the second stage of inflation the potential is simply given by
\be\la{vv}
V(t_2)=U(t_2) -\int_{t_i}^{t}3H\dot{\phi}^2dt=V(t_1) - 3\Delta
H^2+\Delta \rho_a
\ee
where we have used in the last equality that
$U(t_2)=U(t_1)=V(t_1)$. Of course eq.(\ref{vv}) is self-consistent
and it is no surprise since $3\Delta H^2-\Delta \rho_a\simeq
V(t_1) - V(t_2)$ for inflationary epochs where  $E_{k}(t_i) \ll
 V(t_i), i=1,2$. The new point of view is that eq.(\ref{vv})
predicts a second stage of inflation at a much lower scale (seen
from our asymmetric world) and that the value of this scale can be
determined by the scale of the first inflationary period and the
friction term.

\subsection{ A Toy Model}

Let us now study a toy model in the absence of matter or
radiation.  This model is unrealistic for the reasons given at
the end of the section, however, we believe the model illustrates
in a very simple way the main points raised in this work.
We take the ansatz for  the symmetric potential as
$U=A+Bcos^2[\al \phi]$ with $A,B>0$ and $\al$ constants to be
determined. The period for $\phi$ is taken as $0\leq \al\phi\leq
\pi$ with $\phi(t_i=-\infty)=0$ and $\phi(t_f=\infty)=\pi/\al$.
The maxima of the potential are at $\phi(t_i)=0$,
$\alpha\phi(t_f)=\pi$, where $\dot \phi(t_i)=\dot \phi(t_f)=0$,
fixing $A+B\equiv\Lm_1^4$ we can determine $\dot H$ and $H$
giving
\bea
\dot H &=& - \dot\phi^2/2 =-B \sin^2[\al \phi]\\
H&=&\fr{1}{2\sqrt{3}}\,\fr{A(1+\cos[\al\phi])+2B}{\sqrt{A+B}}.
\eea
The integration constant in $H$ is fixed by demanding that at
$t_i$ we have $U(t_i)=V(t_i)=3H^2(t_i)$ since $\dot \phi(t_i)=0$.
The resulting potential is then simply given by $V=3H^2+\dot H$.
By imposing that the minimum of the potential $V$ is zero, i.e.
$V\geq 0$,  and defining the scale at $t_f=\infty$ as
$V(t_f)\equiv \Lm_2^4$ we can determine the constants $A,B,\al$ in
terms of $\Lm_1,\Lm_2$ giving $A=\Lm_1^2(\Lm_1^2-\Lm_2^2),\;
B=\Lm_1^2\Lm_2^2$ and $\al=\sqrt{6}
\Lm_1\Lm_2/(\Lm_1^2-\Lm_2^2)$. In terms of these scales one
has $U=\Lm_1^4(1-(\Lm_2^2/\Lm_1^2)\, \sin^2[\al\phi])$ and
\bea\la{fit}
\phi &=&\fr{1}{\sqrt{6}} \fr{\Lm_1^2-\Lm_2^2}{\Lm_1\Lm_2}\;
\le(2 \arctan[\tanh[\fr{\Lm_1\Lm_2 \al\, (t+t_a)}{\sqrt{2}}]]+\fr{\pi}{2}\ri) \\
\dot\phi  &=&\sqrt{-2\dot H} =\sqrt{2}\Lm_1\Lm_2 \,\sin[\al \phi]    \\
V  &=&\fr{1}{4}\,(\Lm_1^2-\Lm_2^2 +(\Lm_1^2+\Lm_2^2) \cos[\al\phi]\,)^2  \la{Vfi} \\
H  &=&\fr{1}{2\sqrt{3}}\,(\Lm_1^2+\Lm_2^2 +(\Lm_1^2-\Lm_2^2)
\cos[\al\phi]\,).
\eea

In Fig. \ref{fig3} we show the symmetric potential $U(t)$, the
asymmetric potential $V(\phi)$ as well as the acceleration of the
scale factor  of the universe $\ddot a/a=H^2+\dot H$ as functions
of $\phi(t)$. The initial time is taken at the origin
$\phi(t_i=-\infty)=0$ with the maximum of the potential
$V(0)=U(0)=\Lm^4_1$ and  in the limiting solution eq.(\ref{fit})
one has at $t_f=\infty$, $\alpha\phi=\pi$ and $V(t_f)=\Lm_2^4 \ll
U(t_f)=\Lm^4_1$. The acceleration of the universe is positive
around the maxima of the potential. Notice also in Fig.\ref{fig3}
the cyclic nature of the potential $U$ and $V$.

\begin{figure}[htp!]
\begin{center}
\includegraphics[width=7cm]{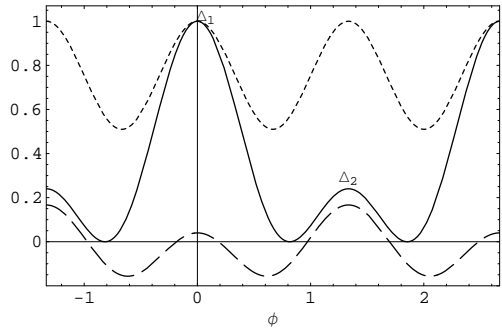}
 \end{center}
\caption{\small{We show $U(t), V(\phi)$ and the acceleration
$\ddot a/a=H^2+\dot H$ as a function of $\phi$ (dotted, solid and
dashed lines, respectively). Notice that $U$ has symmetric maxima
while   $V$ develops a smaller  maximum at $\Lm_2$ with
$\al\phi=\pi$.   Acceleration occurs around the maxima of the
potential.}} \la{fig3}
\end{figure}

In a more realistic situation the inflaton leaves not from the
maximum at $\Lm_1$ but from a slightly displaced position. The
potential is shown in Fig. \ref{fig3}. A mechanism setting the
field away from the maximum at $\phi=0$ is provided by its
fluctuations. We have that
\begin{equation}
\delta \phi \approx \frac{H(t \rightarrow -\infty)}{2\pi}
\approx \frac{\Lambda^2_1}{2\pi\sqrt{3}}.
\label{qf}
\end{equation}

\begin{figure}[htp!]
\begin{center}
\includegraphics[width=6cm]{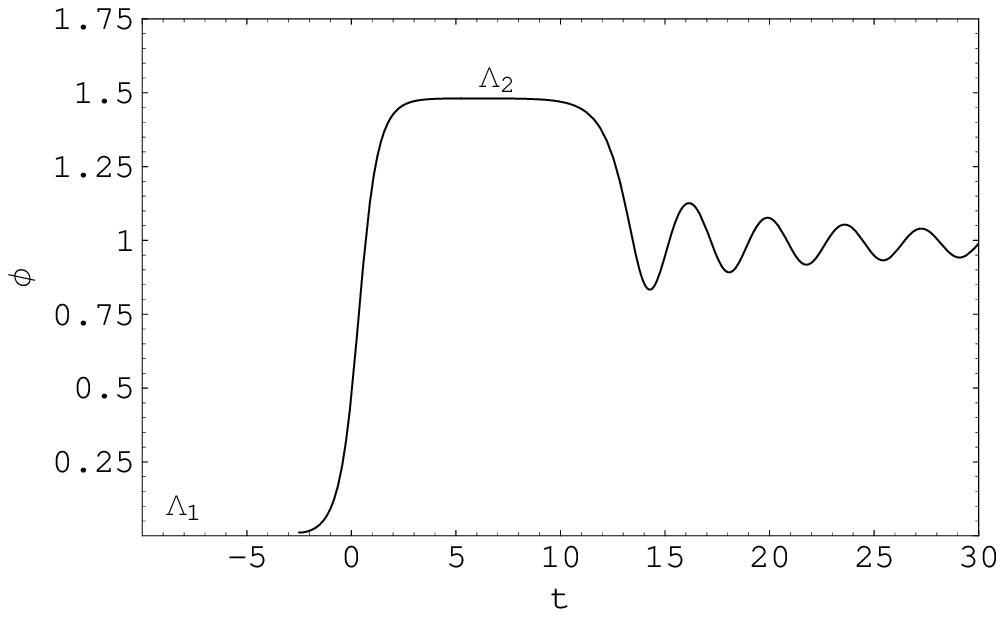}
\includegraphics[width=6cm]{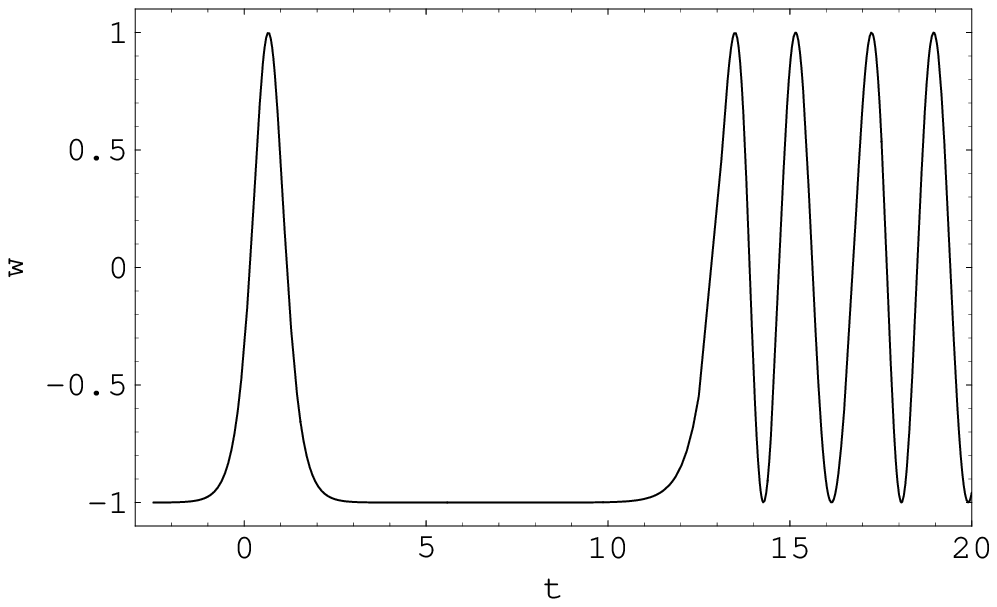}
\end{center}
\caption{\small{The inflaton leaves from close to $\Lm_1$ where it
has been displaced due to its fluctuations $\delta \phi
\approx H(t \rightarrow -\infty)/{2\pi} \approx
\Lambda^2_1/{2\pi\sqrt{3}}$. After some time it approaches the
second maximum at $\Lm_2$  ending in oscillations around the
minimum of the potential.}} \la{fig4}
\end{figure}

Depending on the initial conditions the scalar field approaches
$\Lm_2$ ending  in oscillations around one of the minima. The time
evolution of $\phi$ and state parameter $w=p/\rho$ illustrated in
Fig. \ref{fig4} correspond to a field which is unable to reach the
maximum at $\Lm_2$ ending in oscillations around the first
minimum. For a larger initial kinetic energy the  field would be
able to overcome the maximum at $\Lm_2$ ending at the second
minimum. The beginning and duration of the second inflation depend
on the initial conditions with which the universe was prepared. We
believe this example illustrates in a simple way the main points
raised in this work. However it is not realistic for several
reasons: one can easily show that the end of inflation gives
$H(t_{end})\approx \Lambda^2_2$ which is a very low value for a
realistic $\Lambda_2$. This could be ameliorated by relaxing the
condition that the potential vanishes at the minima. For a
negative potential at the minima there could also be a second
inflation followed by a big crunch. Also there should be a
mechanism of particle production at the end of the first
inflation. This has been originally discussed in terms of
gravitational particle production \cite {ford} and subsequently
criticized \cite {linde} as an inefficient mechanism. A more
efficient one being preheating \cite {linde}, where another scalar
field is used to reheat the universe. It is also possible to
invoke the action of a curvaton field to produce the reheating of
the universe while the inflaton is in charge of inflation only
\cite {li}.

\section{Conclusions}\label{con}

We have studied a model of inflation which can accommodate two
inflationary eras. We use a symmetric potential, valid in a
frictionless world, in which the two inflationary periods have
exactly the same scale. However, we see in our world a second
stage of inflation with a much smaller energy scale as the first
one due to friction terms (expansion of the universe and
interaction with other fields). Both stages of inflation are
derived by the potential energy of a single scalar field. The new
feature is that inflation occurs close to the maxima of the
potential where the kinetic energy is negligible. As a consequence
both inflations are of finite duration. It is then possible that
the second inflation has already or is about to end which should
be testable by substantially increasing the number and accuracy of
supernovae on the Hubble diagram. We show an explicit example
using a toy model. This model is not realistic but nicely
illustrates the main ideas presented in this work. In  the
limiting solution, the scalar field takes an infinite amount of
time to reach the second, smaller, maximum. In a more realistic
case the scalar field is displaced from the higher maximum by its
fluctuations ending in oscillations around a minimum of the
potential. Finally, we have been able to show that a potential of
the type Eq.(\ref{Vfi}) could be derived from supergravity. In
supergravity the only natural scale is the Planck scale and there
are arguments to explain the origin of the first scale of
inflation \cite {ggs} while the second stage of inflation is
derived from the first one by considering friction terms.

\section*{Appendix}

\footnotesize{

Let us consider the supergravity potential for one chiral superfield with
scalar component $z$ and without D-terms~\cite{bailin}
\begin{equation}
 V = e^K
     \left[F^*(K_{zz^*})^{-1}F -
     3|W|^2 \right],
\label{apot0}
\end{equation}
where $
 F \equiv \frac{\partial W}{\partial z} +
       \left(\frac{\partial K}{\partial z}\right) W ,\qquad
 K_{zz^*} \equiv \frac{\partial^2 K}{\partial z \partial z^*}.
$
The reduced Planck mass $M\sim 2.4 \times 10^{18}$ GeV has been set equal
to one. The superpotential and K\"{a}hler potential denoted $W$ and $K$
respectively. Here we are interested in models where $W$ and $K$ are
given by polynomial expressions such as
$ W=\sum_{n=0}^{\infty}a_n z^n, $
and $ K=\sum_{n=1}^{\infty}b_n (zz^*)^n$
where $a_n$ and $b_n$ are real coefficients. In general this structure leads to expressions
that contain $\cos$-form potentials for the angular field $\phi$ which is a real
field defined from $z$ in the following way
$ z=\chi e^{i\phi}$.
By using the above ans\"{a}tze for the superpotential and K\"{a}hler potential,
it is straightforward to show that the supergravity potential can be written
in the form
\begin{equation}
 V = e^K
     \sum_{n=0}^{\infty} \sum_{m=0}^{\infty}\left[\frac{(n+K_1)(m+K_1)}{K_2} -
     3 \right]a_n a_m z^n z^{*m},
\label{apot1}
\end{equation}
where $K_i$ denote the sums
$K_1=\sum_{n=1}^{\infty}nb_n (zz^*)^n$ and $
K_2=\sum_{n=1}^{\infty}n^2b_n (zz^*)^n$.
Let us now insert the radial and angular fields $z=\chi e^{i\phi}$
in eq.(\ref{apot1}).
The  potential is then given by \cite {gaa}
\begin{equation}
 V = e^K
     \sum_{n=0}^{\infty} \sum_{m=0}^{\infty}\left[\frac{(n+K_1)(m+K_1)}{K_2} -
     3 \right]a_na_m\chi^{n+m}\cos[(n-m)\phi],
\label{apot2}
\end{equation}
It is easy to show that Eq.(\ref{apot2}) can give rise to potentials of the type Eq.(\ref{Vfi}).
Let us write the superpotential and K\"{a}ler potential in the form
$W=a_0 + a_1 z + a_2 z^2 $ and $K=z z^* = \chi^2. $
Assuming that the $\chi$ field has relaxed to its v.e.v., $\chi_0$ and eliminating e.g., $a_1$ we
get
\begin{equation}
V(\phi) = c_1 (c_2 + \cos[\phi])^2,
\label{apotex}
\end{equation}
where $ c_1 = 2 e^{\chi_0^2} \chi_0 \sqrt{(\chi_0^2-1) a_0 a_2}$
and
\begin{equation}
c_2 = \frac{((\chi_0^2-2) a_0 + (\chi_0^4+2) a_2)\sqrt{(\chi_0^2-3) a_0^2 - 2 \chi_0^2 (\chi_0^2-1) a_0 a_2 +
\chi_0^2 (\chi_0^4 + \chi_0^2 + 4) a_2^2}}{\sqrt{(\chi_0^2-2)^2 a_0^2 -2 (\chi_0^6-2\chi_0^4+2\chi_0^2+2) a_0 a_2
+(\chi_0^4+2)^2 a_2^2}}.
\label{acedos}
\end{equation}
It is then reasonable that a model of the type Eq.(\ref{Vfi})
could arise from a sugra particle physics model.
}

\section{Acknowledgements}

This work was supported in part by CONACYT project 32415-E and DGAPA, UNAM project IN-110200.

\thebibliography{}

\footnotesize{

\bib{reviews} {For a review and extensive references, see, D. H. Lyth, A. Riotto,
Phys. Rep.314(1999)1-146.
A.R. Liddle, D. Lyth. Comological Inflation and Large Scale Structure, Cambridge
U.P., 2000.}

\bib{papers} {P.J. Peebles and B. Ratra, Astrophys. J. 325(1988)L17.
B. Ratra, and P.J.E. Peebles, Phys. Rev. D37(1988)3406.
C. Wetterich, Nucl. Phys. B302(1988)668.
I. Zlatev, L. Wang and P.J. Steinhardt, Phys. Rev. Lett. 82(1999)896.
R.R. Caldwell, R. Dave and P.J. Steinhardt, Phys. Rev. Lett. 80(1998)1582.
For a recent reviews and further references see, V. Sahni. Class. Quant. Grav.19(2002)3435-3448.
P.J.E. Peebles, B. Ratra, astro-ph/0207347.}

\bib{vilenkin} {P.J.E. Peebles, A. Vilenkin. Phys. Rev. D59(1999)063505.
W.H. Kinney, A. Riotto, Astropart. Phys. 10(1999)387.
M. Peloso, F. Rosati, JHEP 9912(1999)026.
N.J. Nunes, E.J. Copeland, Phys. Rev. D66(2002)043524.
K. Dimopoulos, J.W.F. Valle, Astropart. Phys. 18(2002)287.
For a recent review and further references see, P.J.E. Peebles, B. Ratra, astro-ph/0207347.}

\bib{filippenko} {A. Filippenko, private communication.}

\bib{linde} {G. Felder, L. Kofman, A. Linde. Phys. Rev. D60(1999)103505.
A. H. Campos, H. C. Reis, R. Rosenfeld. hep-ph/0210152}

\bib{ford} {L.H Ford, Phys Rev. D 35(1987)2955; B. Spokoiny, Phys. Lett. B315(1993)40; P.J.E. Peebles
and A. Vilenkin, Phys. Rev. D 59, 0635505 (1999)}

\bib{li} {B. Feng and M. Li, hep-ph/0212213}

\bib{ggs} {G. Germ\'an, G.G Ross, S. Sarkar, Phys. Lett. B469 (1999) 46.
G. Germ\'an, G.G Ross, S. Sarkar, Nucl. Phys. B608 (2001) 423.}

\bib{bailin} {D. Bailin, A. Love, Supersymmetric Gauge Field Theory and String Theory, Hilger, 1994.}

\bib{gaa} {G. Germ\'an, A. Mazumdar, A. Perez-Lorenzana, Mod. Phys. Lett. A17 (2002) 1627.}

\end{document}